\documentclass[showpacs,twocolumn]{revtex4}
\usepackage{epsfig}
\usepackage{graphicx}
\begin{document}
\title{Discussion for the Solutions of Dyson-Schwinger Equations at $m\neq0$ in QED$_3$}
\author{Hui-xia Zhu$^{1,2}$, Hong-tao Feng$^{3,4}$\footnote{Email:fenght@seu.edu.cn}, Wei-min Sun$^{1,4,5}$, and Hong-shi Zong$^{1,4,5}$\footnote{Email:zonghs@chenwang.nju.edu.cn},}
\address{$^{1}$ Department of Physics, Nanjing University, Nanjing 210093, P. R. China}
\address{$^{2}$ The College of Physics and Electronic Information, Anhui  Normal University, Wuhu 241000, China}
\address{$^{3}$ Department of Physics, Southeast University, Nanjing, China}
\address{$^{4}$ State Key Laboratory of Theoretical Physics, Institute of Theoretical Physics, CAS, Beijing 100190, China}
\address{$^{5}$ Joint Center for Particle, Nuclear Physics and Cosmology, Nanjing 210093, China}

\begin{abstract}
In the case of nonzero fermion mass, within a range of $Ans\ddot{a}tze$ for
the full fermion-boson vertex, we show
that Dyson-Schwinger equation for the fermion propagator in QED$_3$
has two qualitatively distinct dynamical chiral symmetry breaking solutions. As the fermion mass increases and reaches to a critical value $m_c$, one solution disappears, and the dependence of $m_c$ on the number of fermion flavors is also given.

\pacs{11.10.Kk, 11.15.Tk, 12.20.-m, 12.40.-y}
\end{abstract}

\maketitle

\section{introduction}

Nowadays, it is widely accepted that Quantum Chromodynamics (QCD) in
3+1 dimensions is the fundamental theory for strong interaction. Dynamical chiral symmetry breaking (DCSB) is of fundamental importance for strong interaction physics. DCSB can be explored via the gap equation, viz., the Dyson-Schwinger equation (DSE) for the dressed-fermion self-energy.
As is well known, the gap equation has two solutions in the chiral limit, i.e. the Nambu-Goldstone (NG) solution which is characterized by DCSB, and the Wigner (WN) solution in which chiral symmetry is not dynamically broken.  
However, when the current quark mass $m$ is nonzero, the quark gap equation has only one solution which corresponds to the NG phase and the solution corresponding to the WN phase does not exist \cite{DSER1,DSER2}. This conclusion is hard to understand and one will naturally ask why the Wigner solution of the quark gap equation only exists in the chiral limit and does not exist at finite current quark mass. The authors of Ref. \cite{zong1} first discussed this problem and asked whether the quark gap equation has a Wigner solution in the case of nonzero current quark mass. Subsequently, the authors
of Refs. \cite{DSE3,DSE4,DSE5,Jiang} further investigated the problem of
possible multi-solutions of the quark gap equation. As far as we
know, partly due to the complexity of the non-Abelian character of QCD, this problem has not been solved satisfactorily in the
literature. In the present paper we try to propose a new approach to
investigate this problem in the framework of a relatively simple Abelian toy model of QCD, namely, quantum electrodynamics in 2+1 dimensions (QED$_3$).

As a field-theoretical model, QED$_3$ has been extensively studied in
recent years. It has many features similar to QCD in 3+1 dimensions.
This is because QED$_3$ is known to have a phase where the 
chiral symmetry of the theory is spontaneously broken and the fermions are confined in this phase \cite{a1}.
Moreover, QED$_3$ is superrenormalizable, so it is not plagued with
the ultraviolet divergences which are present in QED$_4$. These are
the basic reasons why QED$_3$ is regarded as an interesting toy
model: studying QED$_3$ it might be possible to investigate
confinement \cite{a1,a2,a3} and dynamical chiral symmetry
breaking (DCSB) \cite{a4,a5,a6,a7,a8,a9} within a theory which is
structurally much simpler than QCD while sharing the same basic
nonperturbative phenomena. Herein we try to use the DSEs for the
fermion and photon propagators in QED$_3$ to describe novel aspects of the interplay between explicit and dynamical chiral symmetry breaking.

\section{Dyson-Schwinger equation for the fermion propagator}

The Lagrangian of QED$_3$ with $N$ flavors of fermions in a general covariant
gauge in Euclidean space, ignoring the issues discuss in Ref. \cite{a8},
can be written as
\begin{equation}\label{LM}
\mathcal{L}=\sum^N_{j=1}\bar{\psi}_j(\not\!{\partial}+ie\not\!\!{A}-m)\psi_j+\frac{1}{4}F^{2}_{\rho\nu}+\frac{1}{2\xi}(\partial_{\rho}A_{\rho})^2,
\end{equation}
where the 4$\times$1 spinor $\psi_j$ is the fermion field with
$j=1,\cdots, N$ being the flavor indices.

Based on Lorentz structure analysis, the inverse fermion
propagator in the chiral limit can be written as 
\begin{equation}\label{ferp}
    S^{-1}(p)=i\gamma\cdot pA(p^2)+B(p^2).
\end{equation}
One assumes that dressed fermion propagator at finite $m$ ($S^{-1}(m,p)$) is analytic in the neighborhood of $m=0$, so the $S^{-1}(m,p)$ can be written as
\begin{eqnarray}\label{fermm}
&&S^{-1}(m,p)=S^{-1}(p)+\int_0^m \frac{\delta S^{-1}(m',p)}{\delta m'}\mathrm{d}m'\nonumber \\
&&=i\gamma \cdot pA(p^2)+B(p^2)+i\gamma \cdot
pC(m,p^2)+D(m,p^2)\nonumber\\
&&\equiv i\gamma \cdot pE(m,p^2)+F(m,p^2),
\end{eqnarray}
where $E(m,p^2)=A(p^2)+C(m,p^2),~F(m,p^2)=B(p^2)+D(m,p^2)$.

Setting $e^2=1$, the DSE for the fermion propagator can be written as
\begin{eqnarray}\label{eq2}
S^{-1}(m,p)&=&S^{-1}_{0}(m,p)+\int\frac{\mathrm{d}^{3}k}{(2\pi)^{3}}\times\nonumber\\
&&[\gamma_{\rho}S(m,k)\Gamma_{\nu}(m;p,k)D_{\rho\nu}(m,p-k)],
\end{eqnarray}
where $S^{-1}_{0}(m,p)$ is the bare inverse fermion propagator and
$\Gamma_{\nu}(m;p,k)$ is the full fermion-photon vertex. Substituting 
Eq. (\ref{fermm}) into 
Eq. (\ref{eq2}), one can obtain  
\begin{eqnarray}\label{FM1}
&&E(m,p^{2})=1-\frac{1}{4p^2}\int\frac{\mathrm{d}^{3}k}{(2\pi)^{3}}\times\nonumber\\&&~~~~Tr[i(\gamma\cdot
p)\gamma_{\rho}S(m,k)\Gamma_{\nu}(m;p,k)D_{\rho\nu}(m,q)]
\end{eqnarray}
 and 
\begin{eqnarray}\label{FM}
F(m,p^{2})&=&m+\frac{1}{4}\int\frac{\mathrm{d}^{3}k}{(2\pi)^{3}}\times\nonumber\\&&Tr\left[\gamma_{\rho}S(m,k)\Gamma_{\nu}(m;p,k)D_{\rho\nu}(m,q)\right].
\end{eqnarray}
where $q=p-k$.
The full photon propagator can be written as
\begin{equation}
D_{\rho\nu}(m,q)=\frac{\delta_{\rho\nu}-q_{\rho}q_{\nu}/q^{2}}{q^{2}[1+\Pi(m,q^{2})]}+\xi\frac{q_\rho
q_\nu}{q^4},
\end{equation}
with the vacuum polarization $\Pi(m,q^2)$ defined by
\begin{equation}\label{plz}
    \Pi_{\rho\nu}(m,q^2)=(q^2\delta_{\rho\nu}-q_\rho q_\nu)\Pi(m,q^2).
\end{equation}
The DSE satisfied by the photon vacuum polarization tensor reads
\begin{eqnarray}\label{BS}
\Pi_{\rho\nu}(m,q^{2})&=&-N
\int\frac{\mathrm{d}^{3}k}{(2\pi)^{3}}\times\nonumber\\&&Tr\left[S(m,k)\gamma_{\rho}S(m,p)\Gamma_{\nu}(m;q+k,k)\right].
\end{eqnarray}
The boson polarization $\Pi(m,q^2)$ has an ultraviolet divergence
which is present only in the longitudinal part. By
applying the projection operator
\begin{equation}
\mathcal{P}_{\rho\nu}=\delta_{\rho\nu}-3\frac{q_{\rho}q_{\nu}}{q^2},
\end{equation}
one can remove this divergence and obtain a finite vacuum polarization $\Pi(m,q^2)$ \cite{a8}.

The DSEs for the photon and fermion propagators form a set of
coupled integral equations for the three scalar functions $(E(m,p^{2}$, $F(m,p^{2}$ and $\Pi(m,q^2)$) once the full
fermion-photon-vertex $\Gamma(m;p,k)$ is known. Unfortunately, although
several works attempts to resolve the problem, none of them are
completely satisfactory \cite{a10,a11,a12,a13,a14}. Thus, in phenomenological
applications, one often proceed by adopting reasonable approximation for $\Gamma(m;p,k)$ such that Eqs. (\ref{FM1}), (\ref{FM}) and (\ref{BS}) are reduced to a closed system of equations which may be solved directly. In this letter, following Ref. \cite{a8}, we choose the following 
$Ans\ddot{a}tze$ for the full fermion-photon vertex
\begin{equation}
   \Gamma_\nu(m;p,k)=f(E(m,p^2),E(m,k^2))\gamma_\nu,
\end{equation}
and the form of function $\Gamma_\rho(p,k)$ is: (1) $\gamma_\nu$;
(2) $\frac{1}{2}[E(m,p^2)+E(m,k^2)]\gamma_\nu$; (3) $E(m,p^2)E(m,k^2)\gamma_\nu$. The first one is the bare vertex. This
structure plays the most dominant role in the full fermion-photon vertex in high energy region and the full fermion-photon vertex reduces to it in large momentum limit. The second form is inspired by the BC-vertex \cite{a11}. 
Previous works \cite{a8,a16} show that the numerical results of
DSEs employing this $Ans\ddot{a}tze$as is as good as that employing BC and CP vertex \cite{a12}. Since the numerical results obtained using the last $Ans\ddot{a}tze$ coincide very well with earlier investigation \cite{a8,a15}, we choose this one as a reasonable $Ans\ddot{a}tze$ to be used in this work.
Using those $Ans\ddot{a}tze$ for the full fermion-photon vertex, the coupled DSEs for the
fermion propagator and photon vacuum polarization reduce to the
following form\begin{widetext}
\begin{eqnarray}\label{E11}
E(m,p^2)&=&1+\frac{2}{p^2}\int\frac{\mathrm{d}^{3}k}{(2\pi)^3}\frac{E(m,k^2)(p\cdot
q)(k\cdot
q)f(E(m,p^2),F(m,k^2))/q^2}{q^2[E^2(m,k^2)k^2+F^2(m,k^2)][1+\Pi(m,q^2)]},\\
F(m,p^2)&=&m+2\int\frac{\mathrm{d}^{3}k}{(2\pi)^3}\frac{F(m,k^2)f(E(m,p^2),E(m,k^2))}{q^2[E^2(m,k^2)k^2+F^2(m,k^2)][1+\Pi(m,q^2)]},\label{F11}\\
\Pi(m,q^2)&=&2N\int\frac{\mathrm{d}^{3}k}{(2\pi)^3}\frac{2k^2-4(k\cdot
q)-6(k\cdot
q)^2/q^2}{E^2(m,k^2)k^2+F^2(m,k^2)}\frac{E(m,k^2)E(m,p^2)f(E(m,p^2),E(m,k^2))}{q^2[E^2(m,p^2)p^2+F^2(m,p^2)]}\label{P11},
\end{eqnarray}\end{widetext}
where the Landau gauge has been chosen. In the chiral limit, $E(m,p^2)=A(p^2)$ and $F(m,p^2)=B(p^2)$. From Eqs. (12), (13) and (14), it is not difficult to find that the above coupled equations have one Wigner solution
$B(p^2)\equiv0$ and two dynamical symmetry breaking solutions: $B(p^2)$ ($B(p^2)>0$) and
$-B(p^2)$. As was pointed out in Ref. \cite{DSE3}, if $B(p^2)$ is a solution of the gap equation in the chiral limit, then so is $-B(p^2)$. While these two solutions are distinct, the chiral symmetry entails that each yields the same pressure. In the chiral limit, the two dynamical symmetry breaking solutions are symmetric about the Wigner solution $B(p^2)=0$. However, just as will be shown below, this might be changed when the fermion mass is not zero.

\section{numerical results}

\begin{figure}[htp!]
\includegraphics[width=0.39\textwidth]{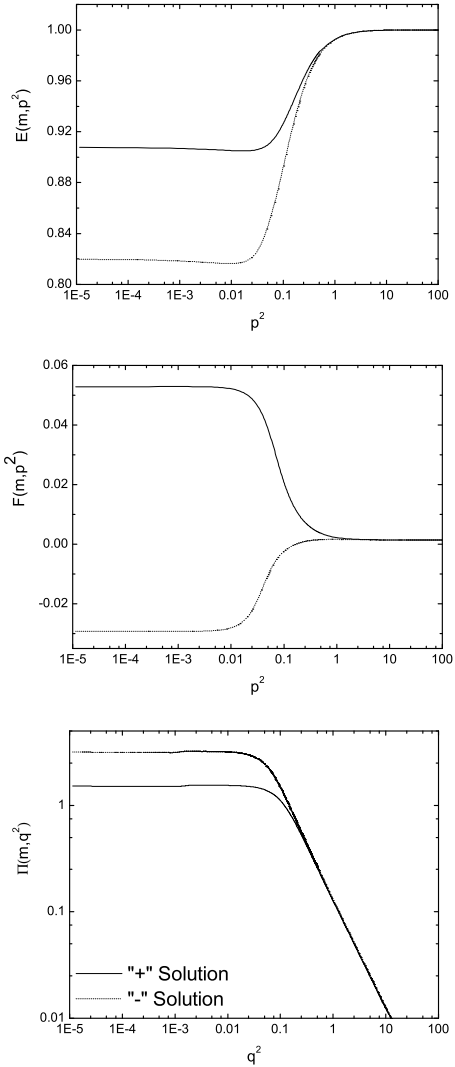}\\
  \caption{The typical behaviors of the two solutions of
DSEs for the fermion propagator at $N=1$, $m=10^{-3}$ for Ansatze
2.}\label{FG1}
\end{figure}

Our next task is to solve for the two scalar functions $E(m,p^2)$ and $F(m,p^2)$. These two functions can be obtained by numerically solving the three coupled integral equations
Eqs. (\ref{E11}-\ref{P11}). Starting from $E=1$, $F=1$ and $\Pi=1$,
we iterate the three coupled integral equations until all the three functions
converge to a stable solution which is plotted in Fig. \ref{FG1} (solid line).

From Fig. 1, it is easy to find that all the three scalar functions in the DCSB phase
($N = 1$) are constant in the infrared region, while in the ultraviolet
region the vector function behaves as $A(p^2)\rightarrow1$ and
the photon vacuum polarization behaves as $\Pi(q^2) \propto 1/q$. Nevertheless,
in contrast to the case of massless
QED$_3$ \cite{a8},
in the large momentum region, the fermion self-energy reduces to the
bare mass $m$ in Eq. (\ref{FM}). Since all the three functions are positive in the whole range of $p^2$, we define them as the "+" solution.

\begin{figure}[htp!]
\includegraphics[width=0.39\textwidth]{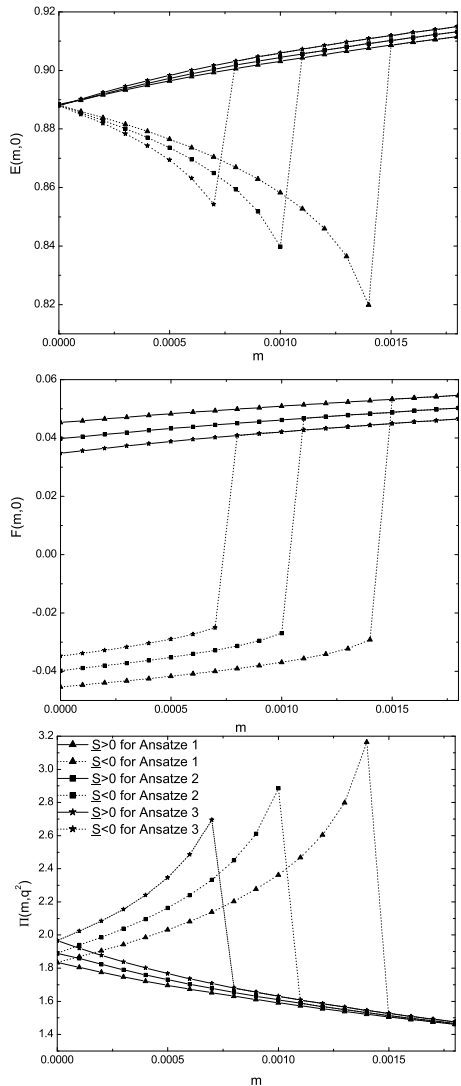}\\
  \caption{The infrared value of $E, F, \Pi$ at N=1 as a function of $m$ ($S>0$ represents the "+" solution and $S<0$ represents the "-" solution).}\label{FG2}
\end{figure}
If we do interation starting from $F=-1,~E=1$ and $\Pi=1$, we can obtain another stable
solution. The typical behaviors of the three functions in the DCSB
phase for a fixed mass and number of fermion flavors are also plotted in
Fig. \ref{FG1} (the dotted line). From Fig. \ref{FG1}, we see that the DSEs for the fermion propagator has two distinct nonzero solutions.
Especially, the infrared value of the fermion self-energy is negative,
so we define it as the "-" solution. In the low energy region, each of the three functions in the second solution is also almost constant, but it is
different from the corresponding one in the "+" solution. As $p^2$ or $q^2$ increases, 
each function of the "-" solution approach to the corresponding one of the "+" solution. 

To reveal the difference between these two solutions, we consider $m$ as
a continuous parameter in the DSEs. We plot the infrared value of
$E,~F,~\Pi$ in Fig. \ref{FG2}. When $m=0$, from DSEs one obtains
one $E$ and $\Pi$, but two $F$ which are symmetric about
$F=0$ in Fig. \ref{FG2} for each vertex ansatze. For the "+" solution of DSEs,
as $m$ increases, $E(0)$ and $F(0)$ increases while $\Pi(0)$
decreases. However, the three infrared values
in the "-" solution show a different trend as $m$ increases.
When $m$ reaches its critical value, we obtain only
one solution for DSEs. In addition, from Fig. \ref{FG2}, it can be seen that the critical mass exists for any truncated scheme of DSEs used in this work.

Furthermore, we investigate the influence of the number of fermion flavors on
the critical mass.
\begin{figure}[htp!]
\includegraphics[width=0.4\textwidth]{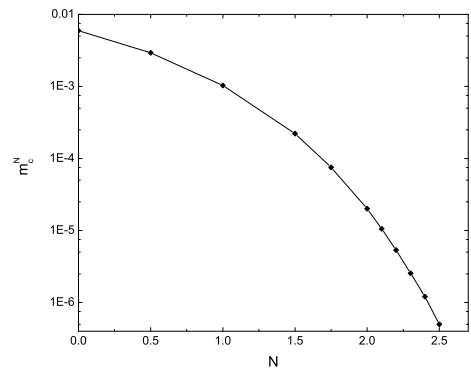}\\[-0.5cm]
\caption{The relation between the critical mass and the number of fermion flavors.}\label{FG3}
\end{figure}
By employing ansatze 2, we can obtain the relation between the
critical mass and the number of fermion flavors and it is plotted in Fig. \ref{FG3}. We observe that the critical mass decreases as $N$ increases and it vanishes at $N=N_{c}$, which is similar to the critical number of fermion flavors for DCSB in the chiral limit \cite{a9}.

\section{conclusions}
To summarize, in this paper, working in the framework of Dyson-Schwinger equations and employing a range of
ansatze for the full fermion-photon vertex of QED$_3$, we study the interplay between explicit and dynamical chiral symmetry breaking in QED$_3$.
In the case of nonzero fermion mass, it is found
that, besides the ordinary solution, the fermions gap equation has another solution which has not been reported in the previous work of QED$_3$.  In
the low energy region, one observes that these two solutions are apparently
different, but in the high energy region they coincide with each other.
In addition, it is found that this solution exists only when the mass is smaller than a critical value. The critical mass decreases apparently with the rise of the number of fermion flavors and vanishes at a critical value $N_c$, which corresponds to the critical number of fermion flavors of QED$_3$ in the chiral limit. It is a interesting phenomena which deserve further investigations.

\section{acknowledgements}
This work was supported by the National Natural Science
Foundation of China (Grant Nos. Nos. 11047005, 11105029,
10935001, and 11075075.) and the Research Fund for the Doctoral Program of Higher Education (under Grant No 2012009111002)

\end{document}